\documentclass[preprint2]{aastex}  

\shorttitle{Catalog of Young Regions Within 1 kpc}
\shortauthors{ Porras et al.}

\begin{document}

\title{A Catalog of Young Stellar Groups and Clusters Within 1 kpc of the Sun}

\author{Alicia Porras, Micol Christopher\altaffilmark{1}, Lori Allen, James Di 
Francesco\altaffilmark{2}, S. Thomas Megeath, and Philip C. Myers}
\affil{Center for Astrophysics, 60 Garden Street, MS-42,
    Cambridge, MA 02138}

\altaffiltext{1}{Currently at Astronomy Department, California Institute 
of Technology, Pasadena, CA 91125}
\altaffiltext{2}{Currently at the Herzberg Institute of Astrophysics, National 
Research Council of Canada. 5071 West Saanich Road, Victoria BC V9E 2E7, 
Canada}

\begin{abstract}
We present a catalog of near-infrared surveys of young ($\la$ a few 10$^6$yr) stellar groups and clusters within 
1 kpc from the Sun, 
based on an extensive search of the literature from the past ten years.  
We find 143 surveys from 69 published articles, covering 73 
different regions. The number distribution of stars in a region has a median of 28 and a mean 
of 100. About 80\% of the stars are in clusters with at least 100 members.
By a rough classification of the groups and clusters based on the number of
their associated stars, we show that most of the stars form in large clusters. 
The spatial distribution of cataloged regions in the Galactic plane shows a relative lack
of observed stellar groups and clusters in the range 270$^\circ< {\it l}<$ 60$^\circ$ of Galactic longitude, reflecting our location between the Local and Sagittarius arms.
This compilation is intended as a useful resource for 
future studies of nearby young regions of multiple star formation.  

\end{abstract}

\keywords{catalogs---infrared:general---open clusters and 
associations:general---stars:formation}

\section{Introduction}

There is much evidence to support the idea that many, and perhaps most, stars 
form in multiple systems (e.g. Lada et al. 1991, Carpenter 2000). 
The degree of observed clustering varies greatly between star-forming 
regions, ranging from the low-density Taurus-Auriga complex, where localized 
stellar densities are typically on order 1-2 pc$^{-3}$ (Jones \& Herbig 1979, 
G\'omez et al. 1993, Larson 1995), to the 
rich Orion Nebula cluster, where the central stellar density 
approaches $2\times10^4$ pc$^{-3}$ (Hillenbrand \& Hartmann 1998).  

How this range of clustering takes place is still under study, and  
clearly there is much to be learned about the processes governing star 
formation through the examination of young multiple systems (Elmegreen et al. 
2000, Meyer et al. 2000). Additionally, in recent years, the evolution in 
infrared (IR) array size and sensitivity has led to a larger sample of 
young multiple systems. 

Therefore, the goal of this paper is to draw together a useful sample of the
closest embedded stellar groups and clusters (within 1 kpc from the Sun), as an aid 
for more detailed comparative studies. This paper extends two recent catalogs:
(1) the catalog of infrared
clusters of Bica, Dutra \& Barbuy (2003), which covers clusters out to distances
$\sim$15 kpc using NIR data as in the 
present study; and (2) the catalog of Lada \& Lada
(2003), which includes clusters within $\sim$2 kpc of the Sun with 
35 or more stars. The present catalog covers a smaller distance from the Sun, but gives 
more entries within that distance than do the catalogs of Bica et al. (2003) and 
Lada \& Lada (2003). 
Many of the regions presented here are also observed in the molecular gas 
($^{13}$CO and C$^{18}$O) survey by Ridge et al. (2003). 

We consider a ``young'' region of multiple star formation to be one where the stars
are still ``embedded'' in or associated with substantial molecular gas. The ages of these stars 
are typically less or about a few Myr according to their positions on the Hertzsprung-Russell diagram 
(Lada \& Lada 2003, Hartmann 2001).

Our sample is limited to 
regions of multiple star formation within 1 kpc since at greater distances, surveys 
begin to suffer significant stellar incompleteness
due to increasing extinction, poor resolution, and poor sensitivity. 
Also, we chose to focus on infrared observations because young groups and clusters 
are often heavily obscured by the gas and dust in their parent molecular 
clouds, and extinction due to interstellar dust is approximately ten times 
less at 2 $\mu$m than in the visual V band. 

Our catalog contains information which we think will be relevant to future 
studies of young stellar groups and clusters. We provide a list of 
regions imaged, cross-indexed to their published surveys. 
We also include areas covered by the surveys, 
their depth, and, where available, completeness, distances to the regions, 
and the number of stars detected.   

For simplicity, in this paper we adopt the term ``regions of multiple star
formation'' or ``regions'' for short, to refer to all 
stellar groupings which meet the criteria of \S 2.1. Then we classify these ``regions'' 
based on the number of their associated stars in \S 3.

\section{Catalog of Regions of Multiple Star Formation}

\subsection{Scope of the Catalog}

In constructing the catalog, 
we wished to compile a resource that would be helpful in future research. 
Consequently, we restricted our catalog to surveys meeting 
certain criteria. 

The survey must:
\begin{itemize}
\item{} contain observations in the near-infrared in at least one of these bands:
J (1.25 $\mu$m), H (1.65 $\mu$m), K (2.2 $\mu$m), K$^{\prime}$ 
(2.11 $\mu$m), or K$_s$ (2.2 $\mu$m). 

\item{} be centered on a young star forming region in association with molecular
gas and/or exhibiting other signposts of star formation, such as the presence of 
Herbig Ae/Be stars, outflows, HII regions or reflection nebulae
at an estimated distance within $\sim$1 kpc of the Sun.

\item{} indicate clustering on at least a minimal scale, considered in this work
to be at least 5 ``associated'' stars.

\item{} have been designed as a survey of an area for near-IR sources 
(as opposed to searches for binaries or other phenomena).

\item{} have been conducted within approximately the last 10 years (up to 
November 2002).

\item{} have been published in any of the following astronomical publications 
searched by ADS: ApJ, ApJS, AJ, MNRAS, PASP, A\&A and A\&AS. 

\end{itemize}

As described above, near-infrared wavelengths were selected because of their 
ability to penetrate the extinction caused by dust in star forming regions. 
Regions within 1 kpc, owing to their distance, have been surveyed to much 
greater depths (i.e. fainter stars have been recorded), allowing for greater 
completeness in the stellar samples. Likewise, developments in observational 
instruments and techniques
have allowed recent surveys to obtain deeper and wider coverage than older surveys, 
while also confirming the information from these previous studies.
Therefore, no 
information is lost by restricting our catalog to the more recent surveys. 
Under these conditions, we expect most large clusters to be included. 

Our surveys catalog contains 143 entries compiled from 69 different papers and 
covering 73 star forming regions. Twenty-five of the entries in our catalog 
were recorded in the Hodapp (1994) survey of IRAS sources with molecular 
outflows and 14 are in recent surveys of Herbig Ae/Be stars by Testi et al. 
(1997; 1998; 1999). Forty-seven of the 73 entries in our catalog of groups and
clusters
also appear in the catalog of Bica, Dutra \& Barbuy (2003) and 25 clusters 
in the catalog of Lada \& Lada (2003).
Twenty-seven of the 30 regions studied by Ridge et al. (2003) are
coincident with our catalog. 

\subsection{Construction of the Catalog}

The first step in constructing the catalog was to identify all regions 
of nearby star formation. This was done through a combination of {\it a priori} 
knowledge and searches on 
the NASA Astrophysical Data System (ADS)\footnote{http://adswww.harvard.edu}.  
Once this list of regions was compiled, ADS searches were conducted on each region, 
searching for articles containing ``infrared'' and the regions' names in the titles. 
For most regions this produced a number of journal articles. Relevant information on 
the surveys described in each of these articles was recorded. In addition, any 
surveys referenced in each article were noted and subsequently investigated. 
This search method produced 109 of the 143 entries recorded in the catalog. 

In addition, SIMBAD\footnote{http://simbad.harvard.edu, 
\mbox{http://simbad.u-strasbg.fr/Simbad}} searches were conducted on these 
regions, either by searching on an approximate central celestial coordinate 
or on stars known within the region. 
In addition, ADS author searches were conducted for authors whose other surveys 
were already included in the catalog. SIMBAD and author searches combined to 
add another 2 entries. 

Finally, this preliminary catalog was sent to 41 scientists involved in near-infrared 
research with requests for their input regarding additional surveys that they had 
conducted, or of which they were aware. Input was received from approximately 
30 people and resulted in 8 entries in our catalog. 

More recently, 24 entries were added based only on ADS searches under ``JHK 
photometry'' and ``young stellar clusters'', bringing the total number 
of surveys to 143, up to November 2002. 
Seventeen of these 24
entries were studies based on the point source catalog of the 2 Micron All 
Sky Survey (2MASS, Beichman et al. 1998) data. In the future, it should be 
possible to extend this catalog based on new data from 2MASS\footnote{available at 
http://www.ipac.caltech.edu/2mass/releases/allsky} and SIRTF observations.

\subsection{Description of the Catalog}

Table~\ref{tbl-1} contains the catalog of near-IR surveys of star-forming regions.
We include in the table all relevant information about each survey that could 
aid in selection for future studies, particularly studies of the spatial 
distribution of stars within clusters.

Because there is considerable overlap between many of the surveys, we present 
in Table~\ref{tbl-2}, a catalog of the groups and clusters themselves. For this reason, columns 1 
and 2 in Table~\ref{tbl-1} contain survey and group or cluster identification numbers. 
These are also included in Table~\ref{tbl-2} to facilitate cross-indexing between 
the tables. 

Columns 3 and 4 in Table~\ref{tbl-1} contain the approximate center position of 
each survey, arranged in order of increasing RA. These centers were either
listed by the authors or were calculated from the RA and Dec coordinates 
given for the entire survey by averaging the minimum and maximum coordinates 
observed. 
Columns 5 and 6 give the spatial extent of each survey, in arc minutes. 
The range denotes the full extent of the survey and not the maximum offset 
from the center position.
Column 7 presents the name of the star-forming region as listed by the author. 

Columns 8 - 10 give the quoted completeness limits of each survey in J, H, and K 
bands respectively. Since completeness estimates are 
reported in various ways, the notes to Table 1 contain explanatory comments.  
Finally, Column 11 contains the references used to compile the catalog.

Some articles, most notably Hodapp (1994) and Testi et al. (1997, 1998, 1999)
contain information on surveys of multiple star-forming regions. 
Each individual region meeting the criteria in \S 2.1 is listed 
separately in the catalog, 
hence the repetition of Hodapp and Testi et al. references. 

Table~\ref{tbl-2} contains the regions covered by the surveys listed in Table~\ref{tbl-1}. 
Columns 1 and 2 are the same as in Table 1 and column 3 gives the most commonly used 
name (or names) for the cluster. Columns 4 and 5 give the cluster RA and Dec. 
In cases where more than one survey was made of a region, 
the coordinates listed in Table~\ref{tbl-2} correspond to the center positions of 
the survey which covered the largest area. We do not attempt to refine the central positions,
however in some cases, more detailed estimates of the coordinates and sizes of regions 
have been done by Bica, Dutra \& Barbuy (2003). Column 6 gives the distance to the 
group or cluster, as given by the authors. In cases where more than one survey has covered the 
region and the distances differ, we have adopted the value more generally accepted. 
Column 7 contains the number of stars ($A$) that are associated with the star-forming
region according to the original authors. When there are several estimates in the literature 
of the number of associated stars in one region, we adopt values that seem to be more complete, 
according to the size of the surveyed area and the sensitivity of the photometric measurements.
In some cases, we adopt the $I_c$ value defined by Testi et al. 
(1999) as ``the integral
over distance of the source surface density profile subtracted by the average 
source density measured at the edge of each field''.
Note that we include only regions with $A\ge$ 5 (\S 2.1).
Finally, the last three columns list the references for the adopted values of coordinates, distance
and associated stars. These numbers are the same as in references at the end of Table~\ref{tbl-1}.

We note that there are some cases in which the values of distance, for groups and clusters in 
the same cloud
or region, may change from author to author. Such is the case of NGC 2023, NGC 2024, NGC 2068 and 
NGC 2071, which are all associated with the Orion B molecular cloud at a distance of $\sim$400-500 pc, 
but values listed in Table~\ref{tbl-2} differ because they correspond to the estimation given by 
different authors. A similar situation occurs with regions XY~Per and LkH$\alpha$~101, and with
regions IRAS~06046-0603, Mon~R2, GGD~12-15, and GGD~17.

\section{Discussion}

\subsection{Classification}

The groups and clusters in Table~\ref{tbl-2} differ greatly in their properties. 
However, a very crude classification based on the number of associated stars in the regions 
is possible, despite the differences in observational sensitivity, 
resolution, and wavelength coverage from region to region.
The number of apparently associated stars is given by the original 
authors for 76 of the 77 (99\%) groups or clusters listed in Table~\ref{tbl-2}. 
We make no attempt 
to improve the quality of these estimates, but take them as given. 

We adopt the term ``region of multiple star formation'' to 
mean a stellar concentration with at least five members.  We use this 
general term to include``clusters'' and ``groups''.  We follow the 
standard usage where a ``cluster'' has a larger spatial extent and/or a 
greater surface density than a ``group''.  The number of members which 
divides a cluster from a group differs somewhat from author to 
author, and here we adopt 30 members, the approximate number of stars 
required for a typical open cluster to survive against evaporation 
(Binney \& Tremaine 1987, Adams \& Myers 2001, Lada \& Lada 2003). For 
convenience we further divide ``small'' and ``large'' clusters at 100 
members.  Thus, we call a region of multiple star formation a ``group'' 
if it has 5-30 members, a ``small cluster'' if it has 31-100 members, 
and a ``large cluster'' if it has more than 100 members.

\subsection{Statistics}
\begin{figure*}[hb]
\plotone{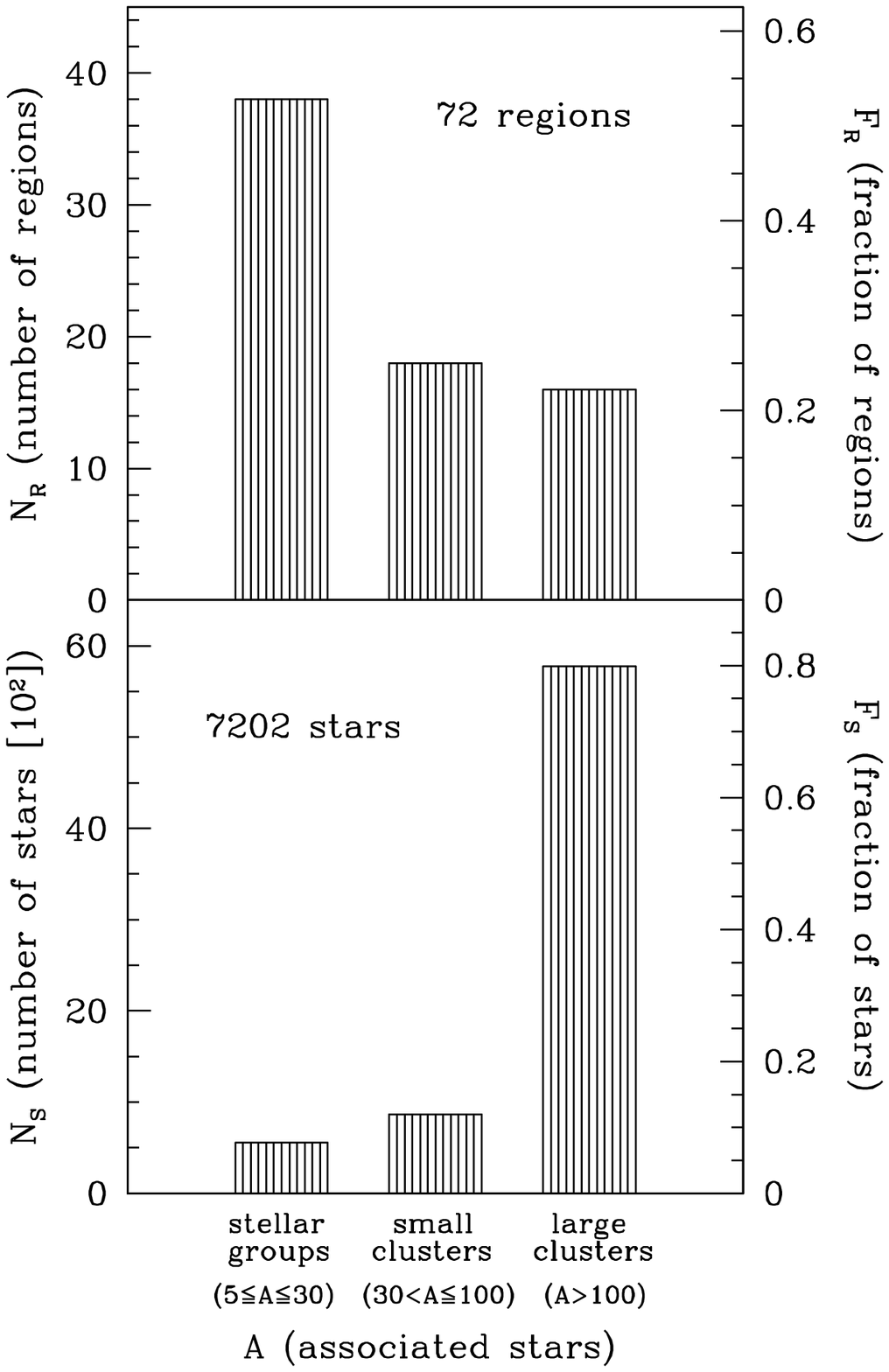}
\caption{ Histogram-bars of the number of regions of multiple star formation (or just ``region'') listed in Table~\ref{tbl-2} (upper) and their associated 
stars (lower). They are classified as ``stellar group'', ``small clusters'', and ``large clusters'' by their number of associated stars. Scales on the right
show the fraction of regions (upper histogram) and the fraction of stars (lower histogram) from the total numbers in the catalog.
\label{fig1}}
\end{figure*}

Considering the number distribution of 7202 associated stars in Table~\ref{tbl-2}, the 
median number of members is 28 and the mean is 100. Nearly all the stars are in the most 
massive clusters: 80\% are in 17 clusters with at least 100 members, and about 50\% are 
in the 5 clusters with at least 345 members. Considering ``groups'', ``small clusters'', 
and ``large clusters'', we note that the choice of dividing lines between these categories 
is arbitrary, but this particular choice gives a substantial number of regions 
(38, 18, and 16) in each category. 
We depict the number of associated stars in a histogram (see Fig.~\ref{fig1}). 
Fig.~\ref{fig1} shows that the number of groups exceeds the number of large clusters, 
but as expected most of the stars are contained in the large clusters. 
In other words, the fraction of the associated stars 
(8\%, 12\%, 80\%) in groups, small and large clusters, trends inversely
with the fraction of these regions (53\%, 25\%, 22\%) in the solar neighborhood.  
This result, reinforces a point made by many investigators (Lada et al. 1991, Carpenter
et al. 2000, Lada \& Lada 2003), that most stars form in large clusters.

\subsection{Spatial Distribution}

\begin{figure*}[ht]
\epsscale{2.3}
\plotone{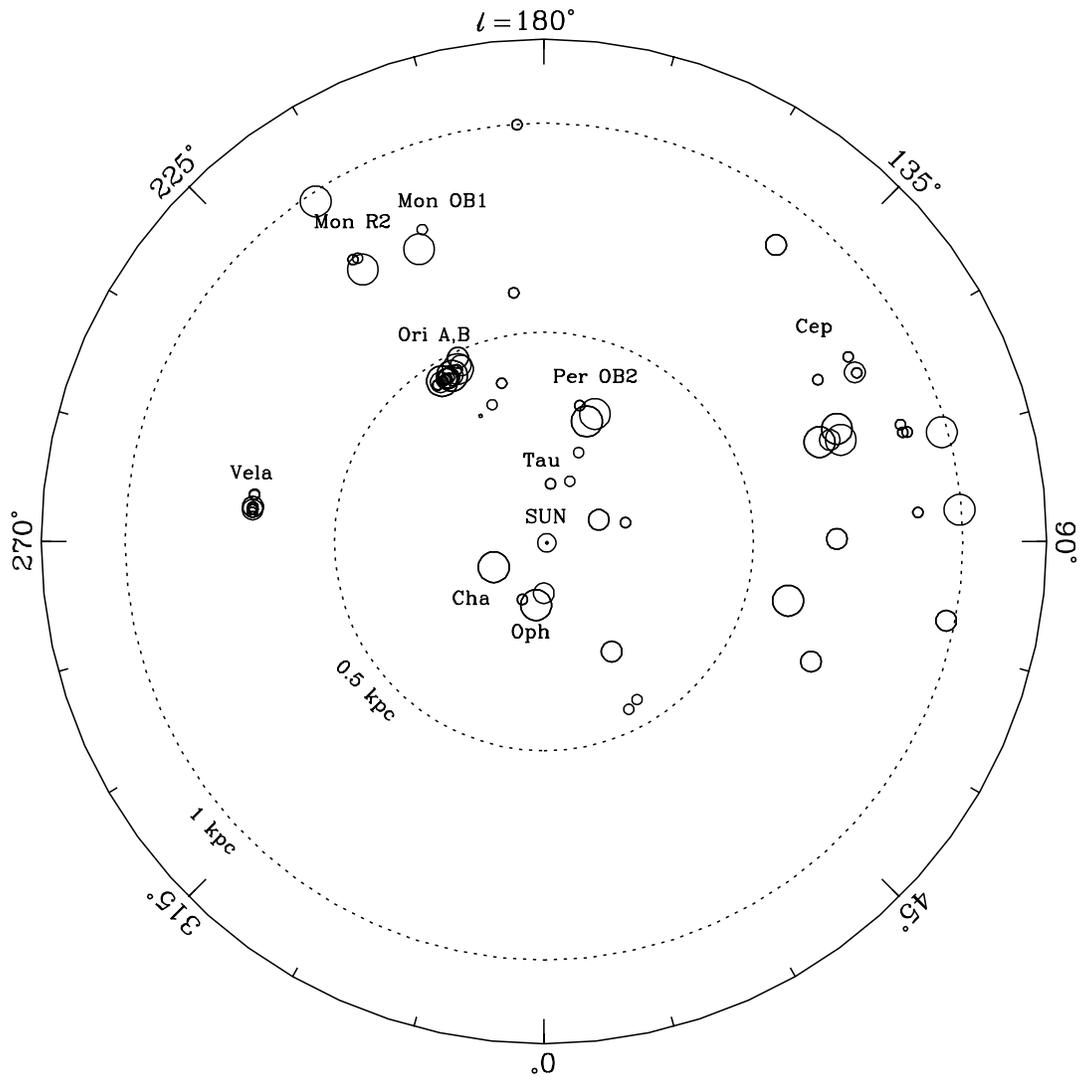}
\caption{ Spatial distribution of young stellar groups and clusters in the Galactic plane. 
Small, medium and large open symbols correspond to A$\leq$30, 30$<$A$\leq$100, and A$>$100 
categories, respectively. Major molecular clouds are labeled.
\label{fig2}}
\end{figure*}

To convey an idea of the Galactic area covered by the surveys in this catalog,
Fig.~\ref{fig2} shows a projected view of the spatial distribution of the young regions. 
A different symbol size is used to show the rough classification of the content of
stars: small circles for
5 $\leq A \leq$ 30, medium size circles for \mbox{30 $< A \leq$ 100}, and large 
circles for $A >$ 100. One dot shows the position of the region without 
estimate of associated stars.
As can be seen, there is not a clear correlation between the number of stars in the 
regions ($A$) and their distance from the Sun. Also, from the K$_{lim}$ and the 
distance values, we conclude that the clusters with more associated stars are not 
preferentially from surveys that are deeper. 

It is striking that the spatial distribution of 
regions of multiple star formation is non-uniform,
with a much lower surface density of observed groups and clusters in the Galactic 
longitude range 270$^\circ< {\it l}<$ 60$^\circ$ than elsewhere. 
This deficiency can be understood by the absence of molecular clouds in the inter-arm 
region between the Local spiral arm and the Sagittarius arm which 
lies towards the forth quadrant at $\sim$1700 pc from the Sun (Dame et al. 1987).

More systematic observations of young clusters, 
for example via 2MASS and SIRTF, will help to greatly 
extend the type of analysis given here, to gain insight into the 
processes which make groups and clusters, as discussed in Meyer et al. 
(2001), Clarke et al. (2000), and Adams \& Myers (2001).

\section{Conclusions}

Our main conclusions are:

\begin{enumerate}

\item Most stars are in a relatively few large clusters.  About 80\% of 
the stars in the sample are in large clusters with more than 100 
members. These large clusters represent 22\% of the regions of 
multiple star formation.

\item  Most regions of multiple star formation are small groups, whose 
total population of stars is relatively small. Groups with 5-30 
members represent 53\% of the regions of multiple star formation, yet 
the total number of stars in such groups is only about 8\% of the 
stars in the sample.

\item The spatial distribution of regions of multiple star formation 
follows roughly the distribution of molecular gas, and shows an 
asymmetry between northern and southern latitudes expected from the 
placement of the nearest Galactic spiral arms.

\end{enumerate}

We thank Tom Dame for helpful discussion, Charles Lada for an advance copy of Lada \& Lada (2003),
and an anonymous referee for useful comments and suggestions.
We thank those who responded to our inquiries regarding additional NIR surveys,
Part of this work was undertaken at the Smithsonian Astrophysical
Observatory Research Experience for Undergraduates (REU) program.
A. P. acknowledges support from the SIRTF Legacy Program via the University
of Texas contract UTA 02-370 to SAO.

\clearpage

\clearpage

\clearpage

\begin{deluxetable}{ccccccccccl}

\tabletypesize{\small}
\rotate
\tablewidth{0pt}
\tablecolumns{13}
\tablecaption{Near-IR Surveys of Young Stellar Groups and Clusters Within 1 kpc \label{tbl-1}}
\tablehead{
\colhead{Survey} & \colhead{Group or} & \colhead{R.A.(2000)} & \colhead{Dec.(2000)}
& \colhead{$\delta$R.A.} & \colhead{$\delta$Dec.}
& \colhead{SFR} & \colhead{$\rm{J_{lim}}$} & \colhead{$\rm{H_{lim}}$} & \colhead{$\rm{K_{lim}}$}
& \colhead{Ref.}
\\
\colhead{\#}  & \colhead{Cluster \#} & \colhead{(center)} & \colhead{(center)}
& \colhead{\arcmin} & \colhead{\arcmin}
& \colhead{} & \colhead{} & \colhead{} & \colhead{}
& \colhead{}
}

\startdata
  1&  1& 00:03:59 &  68:34:41	& 4.9 &6.8&NGC~7822& \nodata & \nodata & 12.3                          &1, 2\\
  2&  2& 00:07:02 &  65:38:39	& 8.5 &8.5&Mac~H12& \nodata & \nodata &17.0                            &3, 4\\
  3&  3& 00:31:30 &  61:58:51	& 8.5 &8.5 &VX~Cas &\nodata &\nodata & 17.0                            &3, 4\\
  4&  4& 00:36:47 &  63:28:59	& 8.5 &8.5 &RNO~1B &\nodata &\nodata & 17.0                            &3, 4\\
  5&  5& 02:56:03 &  20:06:01   &105.0&84.0&MBM~12&16.5 &15.5 &15.0 \tablenotemark{a}                  &52\\
  6&  6& 03:27:39 &  58:46:58	& 8.0 &3.0 &AFGL~490 &\nodata & \nodata & 16.5\tablenotemark{b}        &6\\
  7&  7& 03:28:54 &  31:19:19	&24.0 &30.0 &NGC~1333 &16.5 & 15.5 &14.5                               &7\\
  8&  7& 03:29:04 &  31:16:04   & 8.0 &8.0 &NGC~1333-S/SSV13 &\nodata & \nodata & \nodata              &8\\
  9&  7& 03:29:04 &  31:16:04	&10.0 &10.0 &NGC~1333 &18.6 &16.9 &16.2                                &9\\
 10&  7& 03:29:04 &  31:16:09	& 8.0 &3.0 &HH~7-11/SSV13 & \nodata & \nodata &16.5\tablenotemark{b}   &6\\

 11& 8& 03:44:21 &  32:10:17	&24.0 &24.0 &IC~348 &16.5 &15.5 &14.5                                  &10\\
 12& 8& 03:44:30 &  32:17:00   &40.0 &40.0 &IC~348 &16.5 &15.5 &15.0 \tablenotemark{a}                 &65\\
 13& 8& 03:44:37 &  32:09:19   & 5.0 & 5.0 &IC~348 &15.8 &14.9 &14.3 \tablenotemark{a}                 &69\\
 14& 9& 03:49:36 &  38:58:56	& 7.0 &7.0 &XY~Per & 17.8 & 16.8 & 16.8                                &3, 4, 5\\
 15& 10& 04:30:14 &  35:16:25   &10.7 &4.0 &LkH$\alpha$~101 & 18.2 & 17.3 & 16.8                       &11\\
 16& 11& 04:58:47 &  29:50:37	& 8.5 &8.5&MWC~480&\nodata & \nodata & 17.0                            &3, 4\\
 17& 12& 05:35:09 &  10:01:51	& 7.0 &7.0&HD~245185& 17.5 & 16.2 & 16.8                               &3, 4\\
 18& 13& 05:35:14 & -05:22:24	& 8.0 &3.0&BN/KL& \nodata & \nodata & 16.5\tablenotemark{b}            &6\\
 19& 13& 05:35:16 & -05:23:23	& 0.8 &0.7&Trapezium&\nodata & 15.9 & 16.0\tablenotemark{c}            &13\\

 20& 13& 05:35:16 & -05:23:23	& 6.5&6.5&Trapezium/OMC-1& 18.5 & 17.7 &16.8\tablenotemark{d}                &64\\
 21& 13& 05:35:16 & -05:23:17	&39.0 &39.0&Trapezium & \nodata & \nodata & 14.5\tablenotemark{b}      &14\\
 22& 13& 05:35:16 & -05:23:17	& 1.3 &1.3&Trapezium & \nodata & \nodata & 13.5                        &15\\
 23& 13& 05:35:16 & -05:23:15   &5.1&5.1&Orion Nebula Cluster & \nodata& 17.3 & 17.4\tablenotemark{e}  & 59\\
 24& 13& 05:35:17 & -05:23:23   &5.0&5.0&Trapezium& 18.2 & 17.8 & 17.5                                 & 51\\
 25& 13& 05:35:17 & -05:23:00	&2.3&2.3&Trapezium& \nodata & 17.0 & 18.0                              & 58\\
 26& 13& 05:35:17 & -05:22:00	&2.9&2.4&Trapezium& \nodata & \nodata & 15.0                           & 61\\
 27& 13& 05:35:21 & -05:20:29   & 5.0 & 5.0 &Orion Nebula Cluster&15.8 &14.9 &14.3 \tablenotemark{a}   &69\\
 28& 14& 05:35:28 & -05:11:08	&15.0 &7.0&OMC-2& 14.0 & 14.0 & 14.0\tablenotemark{b}                  &16\\
 29& 14& 05:35:28 & -05:09:40	& 8.0 &3.0&OMC-2& \nodata & \nodata & 16.5\tablenotemark{b}            &6\\

 30& 15& 05:35:29 & -06:26:52	&11.6 &11.6&L1641/HH34/KMS12& 16.8 & 15.8 & 14.7\tablenotemark{c}      &17\\
 31& 16& 05:36:18 & -06:22:10	& 4.0 &4.0&L1641N&\nodata & 17.5 & 17.5\tablenotemark{c}               &18\\
 32& 16& 05:36:19 & -06:22:13	& 8.0 &3.0&L1641N& \nodata & \nodata & 16.5\tablenotemark{b}           &6\\
 33& 16& 05:36:23 & -06:24:50	& 4.0 &4.0&L1641~MSSB 8& \nodata & 17.5 & 17.5\tablenotemark{c}        &18\\
 34& 16& 05:36:23 & -06:23:41	&11.6&11.6&L1641N& 16.8 & 15.8 & 14.7\tablenotemark{c}         &17\\
 35& 16& 05:36:24 & -06:24:56	&8.0&3.0&Morgan 8& \nodata & \nodata & 16.5\tablenotemark{b}           &6\\
 36& 17& 05:36:25 & -06:44:41	&4.0&6.0&L1641/V380~Ori/HH1& \nodata & 17.5 & 17.5 \tablenotemark{c}   &18\\
 37& 17& 05:36:28 & -06:44:46	&11.6&11.6&L1641/V380~Ori& 16.8 & 15.8 & 14.7\tablenotemark{c}         &17\\
 38& 16& 05:36:29 & -06:19:57	&30.0&10.0&L1641N&\nodata & \nodata & 14.5\tablenotemark{f}            &19\\
 39& 18& 05:37:53 & -06:57:09   & 5.0 & 5.0 & V592~Ori &15.8 &14.9 &14.3 \tablenotemark{a}             &69\\

 40& 18& 05:37:54 & -06:57:14	&11.6&11.6&L1641/KMS~35& 16.8 & 15.8 & 14.7\tablenotemark{c}           &17\\
 41& 18& 05:37:55 & -06:56:45	&4.0&4.0&L1641/IRAS~71/T456&\nodata & 17.5 & 17.5\tablenotemark{c}     &18\\
 42& 19& 05:38:42 & -06:58:22	&11.6&11.6&L1641/KMS~36& 16.8 & 15.8 & 14.7\tablenotemark{c}           &17\\
 43& 20& 05:38:45 & -02:36:00   &60.0&60.0&$\sigma$Orionis&16.5 &15.5 &15.0 \tablenotemark{a}          &65\\
 44& 21& 05:38:47 & -07:01:40	&11.6&11.6&L1641C & 16.8 & 15.8 & 14.7\tablenotemark{c}                &17\\
 45& 21& 05:38:47 & -07:01:18	&4.0&4.0&L1641/IRAS~75& \nodata & 17.5 & 17.5\tablenotemark{c}         &18\\
 46& 21& 05:38:47 & -07:01:05	&8.0&3.0&L1641C& \nodata & \nodata & 16.5\tablenotemark{b}             &6\\
 47& 21& 05:38:50 & -07:01:32   & 5.0 & 5.0 &L1641C &15.8 &14.9 &14.3 \tablenotemark{a}                &69\\
 48& 22& 05:39:11 &  04:06:37	&7.0&7.0&HD~37490& 17.7 & 16.6 & 16.6                                  &3, 4, 5\\
 49& 23& 05:39:19 & -07:26:20	&4.0&4.0&L1641/H~4-255& \nodata & 17.5 & 17.5\tablenotemark{c}         &18\\

 50& 24& 05:40:24 &  23:50:54	&8.0&3.0&GGD~4& \nodata & \nodata & 16.5\tablenotemark{b}              &6\\
 51& 25& 05:40:47 & -08:06:15	&11.6&11.6&L1641/CK~Group& 16.8 & 15.8 & 14.7\tablenotemark{c}         &17\\
 52& 25& 05:40:49 & -08:06:51	&8.0&3.0&L1641~S4& \nodata & \nodata & 16.5\tablenotemark{b}           &6\\
 53& 25& 05:40:49 & -08:05:58	&6.0&6.0&L1641/IRAS 193,194& \nodata & 17.5 & 17.5\tablenotemark{c}    &18\\
 54& 27& 05:41:36 & -01:55:04	&7.0&5.5&NGC~2024& 19.5 & 18.5 & 17.0                                  &20\\
 55& 27& 05:41:37 & -01:53:39   & 5.0 & 5.0 &NGC~2024 &15.8 &14.9 &14.3 \tablenotemark{a}              &69\\
 56& 26& 05:41:44 & -02:14:31	&3.0&2.0&NGC~2023& 15.0 & 15.0 & 15.0                                  &21\\
 57& 27& 05:41:45 & -01:54:32	&11.5&13.3&NGC~2024& 16.5&15.5&14.5\tablenotemark{b}                   &60\\

 58& 27& 05:41:49 & -01:55:17	&8.0&3.0&NGC~2024&\nodata & \nodata & 16.5\tablenotemark{b}            &6\\
 59& 26, 27& 05:42:16 & -02:19:19	&50.0&35.0&L1630& 16.5 & 15.5 & 14.5\tablenotemark{b}          &22\\
 60& 28& 05:42:27 & -08:09:12   & 5.0 & 5.0 &L1641S &15.8 &14.9 &14.3 \tablenotemark{a}                &69\\
 61& 28& 05:42:29 & -08:06:47	&11.6&11.6&L1641S&16.8 & 15.8 & 14.7\tablenotemark{c}                  &17\\
 62& 29& 05:42:38 & -10:00:51   & 5.0 & 5.0 &IRAS~05401-1002 &15.8 &14.9 &14.3 \tablenotemark{a}       &69\\
 63& 30& 05:42:48 & -08:39:17	&4.0&6.0&L1641/IRAS~243,245& \nodata & 17.5 & 17.5\tablenotemark{c}    &18\\
 64& 30& 05:42:50 & -08:38:37   & 5.0 & 5.0 &IRAS~05404-0839 &15.8 &14.9 &14.3 \tablenotemark{a}       &69\\
 65& 31& 05:42:54 & -09:45:38   & 5.0 & 5.0 &IRAS~05404-0946 &15.8 &14.9 &14.3 \tablenotemark{a}       &69\\
 66&26, 27, 32, 34& 05:43:44 & -01:04:25&140.0&200.0&L1630& \nodata & \nodata & 13.0\tablenotemark{g}  &23\\

 67& 32& 05:46:47 &  00:06:16   & 5.0 & 5.0 &NGC~2068 &15.8 &14.9 &14.3 \tablenotemark{a}              &69\\
 68& 33& 05:47:01 &  21:00:25	&4.5&4.5&CB~34& 14.5 & 14.5 & 14.5                                     &24\\
 69& 34& 05:47:04 &  00:21:44	&8.0&3.0&NGC~2071&\nodata & \nodata & 16.5\tablenotemark{b}            &6\\
 70& 35& 06:07:08 & -06:03:53   & 5.0 & 5.0 &IRAS~06046-0603 &15.8 &14.9 &14.3 \tablenotemark{a}       &69\\
 71& 36& 06:07:45 & -06:22:53	&4.0&4.0&Mon~R2& \nodata & 17.4 & 16.2\tablenotemark{c}                &25\\
 72& 36& 06:07:46 & -06:23:07	&1.7&1.7&Mon~R2&\nodata & \nodata & 16.0                               &26\\
 73& 36& 06:07:46 & -06:23:07	&1.3&1.3&Mon~R2&\nodata & 19.2 & 18.3\tablenotemark{c}                 &27\\
 74& 36& 06:07:46 & -06:23:00	&8.0&3.0&Mon~R2& \nodata & \nodata & 16.5\tablenotemark{b}             &6\\
 75& 36& 06:07:46 & -06:22:29	&15.0&15.0&Mon~R2& 16.2 & 14.6 & 13.4\tablenotemark{c}                 &28\\
 76& 36& 06:07:47 & -06:22:29	&3.2&3.2&Mon~R2& 18.2 & 17.6 & 16.6\tablenotemark{c}                   &28\\

 77& 36& 06:07:48 & -06:22:20   & 5.0 & 5.0 &Mon~R2 &15.8 &14.9 &14.3 \tablenotemark{a}                &69\\
 78& 37& 06:10:49 & -06:11:38   & 5.0 & 5.0 &GGD~12-15 &15.8 &14.9 &14.3 \tablenotemark{a}             &69\\
 79& 37& 06:10:51 & -06:11:54	&8.0&3.0&GGD~12-15& \nodata & \nodata & 16.5\tablenotemark{b}          &6\\
 80& 38& 06:12:48 & -06:13:56   & 5.0 & 5.0 &GGD~17 &15.8 &14.9 &14.3 \tablenotemark{a}                &69\\
 81& 39& 06:31:07 &  10:26:05	&7.0&7.0&VY~Mon& 17.7 & 16.9 & 16.4                                    &3, 4\\
 82& 40& 06:40:35 &  09:37:09	&16.0&31.0&NGC~2264&14.5 & 14.5 & 14.5                                 &29\\
 83& 40& 06:40:45 &  09:48:02	&7.0&7.0&LkH$\alpha$~25& 17.9 & 16.0 & 16.0                            &3, 4, 5\\
 84& 40& 06:41:03 &  09:36:10	&8.0&3.0&Mon~OB1-D& \nodata & \nodata & 16.5\tablenotemark{b}          &6\\
 85& 40& 06:41:05 &  09:40:48   &45.0&45.0&NGC~2264&14.7&14.1&13.9                                     &66\\
 86& 40& 06:41:10 &  09:29:36	&0.1&0.1&NGC~2264 IRS& \nodata & \nodata & \nodata                     &30\\

 87& 40& 06:41:11 &  09:29:32	&8.0&3.0&NGC~2264& \nodata & \nodata & 16.5\tablenotemark{b}           &6\\
 88& 41& 08:39:21 & -41:19:53	&2.0&2.0&Vela/IRAS~08375-4109& 15.5 & 15.5 & 15.5\tablenotemark{d}     &32, 33\\
 89& 42& 08:41:15 & -40:52:17	&2.0&2.0&Vela/IRAS~08393-4041& 15.5 & 15.5 & 15.5\tablenotemark{d}     &31, 32\\
 90& 43& 08:42:06 & -40:44:09	&2.0&2.0&Vela/IRAS~08404-4033& 15.5 & 15.5 & 15.5\tablenotemark{d}     &32, 33\\
 91& 44& 08:46:35 & -43:54:30	&2.0&2.0&Vela/IRAS~08448-4343& 15.5 & 15.5 & 15.5\tablenotemark{d}     &32, 33\\
 92& 45& 08:48:48 & -42:54:29	&2.0&2.0&Vela/IRAS~08470-4243& 15.5 & 15.5 & 15.5\tablenotemark{d}     &32, 33\\
 93& 46& 08:48:48 & -43:32:29	&2.0&2.0&Vela/IRAS~08470-4321& 15.5 & 15.5 & 15.5\tablenotemark{d}     &32, 33\\
 94& 47& 08:49:26 & -43:17:11	&2.0&2.0&Vela/IRAS~08476-4306& 15.5 & 15.5 & 15.5\tablenotemark{d}     &32, 33\\
 95& 48& 08:49:33 & -44:10:45	&2.0&2.0&Vela/IRAS~08477-4359& 15.5 & 15.5 & 15.5\tablenotemark{d}     &32, 33\\

 96& 49& 08:52:18 & -43:05:40	&2.0&2.0&Vela/IRAS~08500-4254& 15.5 & 15.5 & 15.5\tablenotemark{d}     &31, 32\\
 97& 50& 08:53:09 & -42:13:03	&3.0&3.0&Vela Mol. Ridge/BBW192E& \nodata &15.3 & 14.4\tablenotemark{c}&57\\
 98& 51& 11:03:30 & -78:02:23	&87.0&176.0&Cha~I& 16.0 & \nodata & 14.0                               &34\\
 99& 51& 11:06:00 & -77:30:00	&87.0&176.0&Cha~I& 16.0 & \nodata & 13.5                               &35\\
100& 51& 11:07:26 & -77:36:49	&10.0&10.5&Cha~I& 18.0 & 17.0 & 16.5\                                  &36\\
101& 51& 11:09:30 & -76:34:44	&2.1&2.1&Cha~INc& 19.2& 18.9& 18.2\tablenotemark{h}                    &62\\
102& 51& 11:09:45 & -76:33:29	&2.8&2.8&Cha~INa& 19.2& 18.9& 18.2\tablenotemark{h}                    &62\\
103& 51& 11:09:55 & -76:34:18	&5.5&5.5&Cha~I& 18.1 & 17.0 & 16.2\tablenotemark{d}                    &55\\
104& 51& 11:10:03 & -76:37:33	&2.8&2.8&Cha~INb& 19.2& 18.9& 18.2\tablenotemark{h}                    &62\\
105& 51& 11:10:30 & -77:30:00	&39.0&90.0& Cha~I & 16.5&16.0&15.0                                     &63\\
106& 52& 16:09:13 & -39:04:52	&7.0&11.0&Lup~3& 17.2 & 16.5 & 15.8\tablenotemark{b}                   &56\\

107& 53& 16:26:16 & -24:29:15	&98.0&45.0&Oph & 15.5 & 14.5 & 14.0\tablenotemark{f}                   &37\\
108& 53& 16:26:24 & -24:22:35   &8.75&6.25&L1688 core A& 21.5& 21.0&\nodata                            &68\\
109& 53& 16:26:31 & -24:22:59	&90.0&45.0&Oph & 15.0 & 14.5 & 14.0\tablenotemark{d}                   &38\\
110& 53& 16:26:52 & -24:31:42	&100.0&40.0&Oph & 17.0 & 17.0 & 15.5                                   &39\\
111& 53& 16:26:52 & -24:29:12	&30.0&25.0&Oph & 16.5 & 15.4 & 14.2\tablenotemark{c}                   &40\\
112& 53& 16:27:02 & -24:36:20   & 6.5& 7.5&L1688 core E/F& 21.5& 21.0&\nodata                          &68\\
113& 53& 16:27:12 & -24:34:41	&13.6&10.0&Oph & \nodata & \nodata & 14.2                              &41\\
114& 53& 16:27:16 & -24:33:19	&26.8&26.8&Oph & 16.5 & 14.5 & 13.0\tablenotemark{b}                   &42\\
115& 53& 16:27:23 & -24:27:40   & 6.5& 7.5&L1688 core B& 21.5& 21.0&\nodata                            &68\\
116& 54& 18:27:39 & -03:49:52	&8.5&8.5&MWC~297& \nodata & \nodata & 16.7                             &3, 4\\

117& 55& 18:28:48 &  00:08:40	&8.5&8.5&VV~Ser& \nodata & \nodata & 16.9                              &3, 4\\
118& 56& 18:29:54 &  01:14:35	&6.0&5.0&Serpens& 18.0 & 16.0 & 15.5                                   &43\\
119& 56& 18:29:54 &  01:14:35	&4.0&5.0&Serpens& 18.4 & 17.9 & 16.3                                   &44\\
120& 56& 18:29:55 &  01:14:35	&7.0&15.0&Serpens& 15.5 & 14.8 & 14.0\tablenotemark{f}                 &45\\
121& 56& 18:29:57 &  01:14:46	&8.0&3.0&Serpens SVS2&\nodata & \nodata & 16.5\tablenotemark{b}        &6\\
122& 56& 18:29:59 &  01:14:25   & 8.0 & 6.0 &Serpens (SVS2,SVS4,SVS20) &18.0 &18.5 &17.5               &12\\
123& 57& 19:01:54 & -36:57:10	&8.0&3.0&R~Cr~A&\nodata & \nodata & 16.5\tablenotemark{b}              &6\\
124& 57& 19:02:01 & -36:59:09	&18.8&15.0&Corona Australis& 17.5 & 17.0 & 16.5\tablenotemark{b}       &47\\
125& 57& 19:09:23 & -37:25:08	&72.0&40.0&Corona Australis& 14.0 & 14.0 & 15.0\tablenotemark{c}       &46\\
126& 58& 20:07:06 &  27:28:59	&5.0&5.0&IRAS~20050+2720& 17.8 & 17.5 & 16.5\tablenotemark{c}          &48\\

127& 59& 20:20:28 &  41:21:51	&2.2&2.3&BD~+40\arcdeg4124& 18.8 & 17.8 & 15.3\tablenotemark{d}        &49\\
128& 59& 20:20:29 &  41:21:51	&7.0&7.0&BD~+40\arcdeg4124& 17.3 & 16.1 & 16.0                         &3, 4, 5\\
129& 60& 20:27:27 &  37:22:48	&7.0&4.0&S~106& \nodata & \nodata & 14.0                               &50\\
130& 61& 20:57:14 &  77:35:47	&8.0&3.0&L1228&\nodata & \nodata & 16.5\tablenotemark{b}               &6\\
131& 62& 21:03:58 &  50:14:38	&8.0&3.0&L988-e&\nodata & \nodata & 16.5\tablenotemark{b}              &6\\
132& 63& 21:42:50 &  66:06:36	&8.5&8.5&BD~+65\arcdeg1637& \nodata & \nodata & 17.0                   &3, 4\\
133& 63& 21:43:02 &  66:06:29	&8.0&3.0&LkH$\alpha$~234&\nodata & \nodata & 16.5\tablenotemark{b}     &6\\
134& 64& 21:53:29 &  47:16:00   &6.4&6.4&IC~5146& \nodata & \nodata & 16.5                             &67\\
135& 65& 21:54:19 &  47:12:11	&8.5&8.5&LkH$\alpha$~257&\nodata & \nodata & 17.0                      &3, 4\\
136& 66& 22:06:49 &  59:02:11	&2.0&2.0&Gy~2-21& \nodata & \nodata & \nodata                          &53\\

137& 67& 22:19:18 &  63:18:48	&1.3&1.3&S~140 & \nodata & \nodata & \nodata                           &54\\
138& 68& 22:19:28 &  63:32:56	&8.0&3.0&S~140N&\nodata & \nodata & 16.5\tablenotemark{b}              &6\\
139& 69& 22:28:52 &  64:13:44	&8.0&3.0&L1206&\nodata & \nodata & 16.5\tablenotemark{b}               &6\\
140& 70& 22:47:17 &  62:01:58	&8.0&3.0&L1211&\nodata & \nodata & 16.5\tablenotemark{b}               &6\\
141& 71& 22:53:15 &  62:08:45	&8.5&8.5&HD~216629& \nodata & \nodata & 17.0                           &3, 4\\
142& 72& 22:56:19 &  62:01:58	&8.0&3.0&Cep~A& \nodata & \nodata & 16.5\tablenotemark{b}              &6\\
143& 73& 23:05:49 &  62:30:02	&8.0&3.0&Cep~C& \nodata & \nodata & 16.5\tablenotemark{b}              &6\\

\enddata
\tablecomments{{\scriptsize Units of right ascension are hours, minutes, and seconds, and units of declination are degrees, arcminutes, and arcseconds.}}
\tablenotetext{a}{from 2MASS data}
\tablenotetext{b}{90\% completeness limit.}
\tablenotetext{c}{5 $\sigma$ detection limit.}
\tablenotetext{d}{10 $\sigma$ detection limit.}
\tablenotetext{e}{75\% completeness limit.}
\tablenotetext{f}{95\% completeness limit.}
\tablenotetext{g}{92\% completeness limit.}
\tablenotetext{h}{3 $\sigma$ detection limit.}

\tablerefs{{\scriptsize (1) Sugitani et al. 1995; (2) Sugitani et al. 1991; (3) Testi et al. 1998; (4) Testi 
et al. 1999; (5) Testi et al. 1997; (6) Hodapp 1994; (7) Lada et al. 1996; (8) Aspin \& Sandell 
1997; (9) Aspin et al. 1994; (10) Lada \& Lada 1995; (11) Aspin \& Barsony 1994; (12) Kaas 1999; 
(13) Petr et al. 1998; (14) Ali \& DePoy 1995; (15) McCaughrean \& Stauffer 1994; (16) 
Jones et al. 1994; (17) Strom et al. 1993; (18) Chen \& Tokunaga 1994; (19) Hodapp \& Deane 1993; 
(20) Comeron et al. 1996; (21) DePoy et al. 1990; (22) Li et al. 1997; (23) Lada et al. 1991; 
(24) Alves \& Yun 1995; (25) Yao et al. 1997; (26) Aspin \& Walther 1990; (27) Howard et al. 1994; 
(28) Carpenter et al. 1997; (29) Lada et al. 1993; (30) Thompson et al. 1998; (31) Lorenzetti et 
al. 1993; (32) Massi et al. 1999; (33) Liseau et al. 1992; (34) Cambresy et al. 1997; (35) Cambresy 
et al. 1998; (36) Comeron et al. 1999; (37) Kenyon et al. 1998; (38) Barsony et al. 1997; 
(39) Comeron et al. 1993; (40) Strom et al. 1995; (41) Barsony et al. 1989; (42) Greene \& Young 
1992; (43) Eiroa \& Casali 1992; (44) Giovannetti et al. 1998; (45) Sogawa et al. 1997; (46) 
Wilking et al. 1992; (47) Wilking et al. 1997; (48) Chen et al. 1997; (49) Hillenbrand et al. 
1995; (50) Hodapp \& Rayner 1991; (51) Muench et al. 2002; (52) Luhman 2001; (53) Tapia et al. 
1997; (54) Evans et al. 1989; (55) Oasa et al. 1999; (56) Nakajima et al. 2000; (57) Burkert et al. 
2000; (58) Luhman et al. 2000; (59) Hillenbrand \& Carpenter 2000; (60) Haisch et al. 2000; 
(61) Simon et al. 1999; (62) Persi et al. 1999; (63) G\'omez et al. 2000; (64) Lada et al. 2000;
(65) Tej et al. 2002; (66) Rebull et al. 2002; (67) Herbig \& Dahm 2002; (68) Allen et al. 2002; 
(69) Carpenter 2000; (70) Massi et al. 2000; (71) Tuthill et al. 2002.}}

\end{deluxetable}

\clearpage

\begin{deluxetable}{cccccccccc}

\tabletypesize{\footnotesize}
\rotate
\tablewidth{0pt}
\tablecolumns{10}
\tablecaption{Young Stellar Groups and Clusters within 1 kpc of the Sun \label{tbl-2}}
\tablehead{
\colhead{Group or} & \colhead{Survey \#} & \colhead{Cluster Name} & \colhead{RA} 
& \colhead{Dec} & \colhead {Distance} & \colhead {Associated} 
& \multicolumn{3}{c}{References for adopted values} \\
\colhead{Cluster \#} & \colhead{} & \colhead{} & \colhead{(2000)} & \colhead{(2000)} & \colhead{(pc)} 
& \colhead{stars} & \colhead {Coordinates} & \colhead {Distance} 
& \colhead {Associated stars}} 
\startdata

 1&1&NGC~7822&                                        	00:03:59 &  68:34:41  	   &850 	& 40	& 2 & 1 & 1 \\
 2&2&Mac~H12&                                         	00:07:02 &  65:38:39  	   &850 	& 5$\pm$1 & 3 & 3 & 4 \\
 3&3&VX~Cas&                                          	00:31:30 &  61:58:51  	   &760 	& 5$\pm$4 & 3 & 3 & 4 \\
 4&4&RNO~1B&                                          	00:36:47 &  63:28:59  	   &850 	&10$\pm$1 & 3 & 3 & 4 \\
 5&5&MBM~12&                                            02:56:03 &  20:06:01       &275 	& 12 & 52 & 52 & 52 \\
 6&6&AFGL~490&                                        	03:27:39 &  58:46:58  	   &900 	& 45 & 6 & 6 & 6 \\
 7&7 to 10&NGC~1333&                                    03:28:54 &  31:19:19  	   &350         & 143 & 7 & 6,9 & 7 \\
 8&11, 12, 13&IC~348&                                   03:44:30 &  32:17:00  	   &320		&345 & 65 & 10 & 10 \\
 9&14&XY~Per&                                        	03:49:36 &  38:58:56  	   &160 	&11$\pm$3 &  3 & 3 & 4 \\

10&15&LkH$\alpha$~101&                               	04:30:14 &  35:16:25  	   &340         &16 &  11 & 71 & 11 \\
11&16&MWC~480&                                       	04:58:47 &  29:50:37  	   &140 	&5$\pm$6 &  3 & 3 & 4 \\
12&17&HD~245185&                                     	05:35:09 &  10:01:51  	   &400 	&5$\pm$5 &  3 & 3 & 4 \\
13&18 to 27&Trapezium/ONC&                              05:35:16 & -05:23:17  	   &480 	& 1471 &  14 & 14 & 14 \\
14&28, 29&OMC-2&                                     	05:35:28 & -05:11:08  	   &480 	& 33 &  16 & 14 & 14 \\
15&30&L1641/KMS~12&                                  	05:35:29 & -06:26:52  	   &480 	& 5 & 17 & 17 & 17 \\
16&31 to 35, 38&L1641N&                                 05:36:23 & -06:23:41  	   &480 	&43 & 17 & 17 & 17 \\
17&36, 37&L1641/V380~Ori&                        	05:36:28 & -06:44:46  	   &480 	& 34 & 17 & 17 & 17 \\
18&39, 40, 41&L1641/KMS~35&                             05:37:54 & -06:57:14  	   &480 	&15 & 17 & 17 & 17 \\
19&42&L1641/KMS~36&                                  	05:38:42 & -06:58:22  	   &480 	&10 & 17 & 17 & 17 \\

20&43&$\sigma$Orionis&                                  05:38:45 & -02:36:00       &352 	&\nodata & 65 & 65 & 65 \\
21&44 to 47&L1641C&                                     05:38:47 & -07:01:40  	   &480 	&47 & 17 & 17 & 17 \\
22&48&HD~37490&                                      	05:39:11 &  04:06:37  	   &360 	&10$\pm$3 & 5 & 5 & 4 \\
23&49&L1641/H~4-255 &                                	05:39:19 & -07:26:20  	   &480 	& 6 & 18 & 18 & 18 \\
24&50&GGD~4&                                         	05:40:24 &  23:50:54  	   &1000 	&10 & 6 & 6 & 6 \\
25&51, 52, 53&L1641/CK~Group&                           05:40:47 & -08:06:15  	   &480 	&25 & 17 & 17 & 17 \\
26&56, 59, 66&NGC~2023&                              	05:41:44 & -02:14:31  	   &480 	& 16 & 21 & 21 & 21 \\
27&54, 55, 57, 58, 59, 66&NGC~2024&                     05:41:45 & -01:54:32  	   &415 	& 257 & 60 & 60 & 60 \\

28&60, 61&L1641~S&                                      05:42:29 & -08:06:47  	   &480 	&134& 17 & 17 & 17 \\
29&62&IRAS~05401-1002&                                  05:42:38 & -10:00:51       &480         &22 & 69 & 69 & 69 \\
30&63, 64&L1641/IRAS~243, 245&                          05:42:48 & -08:39:17  	   &480 	&40 & 18 & 18 & 18 \\
31&65&IRAS~05404-0946&                                  05:42:54 & -09:45:38       &480         &23 & 69 & 69 & 69  \\
32&66, 67&NGC~2068&                                     05:46:47 &  00:06:16       &480         &45 & 69 & 69 & 69  \\
33&68&CB~34&                                         	05:47:01 &  21:00:25  	   &600 	&12 & 24 & 24 & 24 \\
34&66, 69&NGC~2071&                                  	05:47:04 &  00:21:44  	   &500 	& 97 & 6 & 6 & 6 \\
35&70&IRAS~06046-0603&                                  06:07:08 & -06:03:53       &830         &15 & 69 & 69 & 69 \\
36&71 to 77&Mon~R2&                   	                06:07:46 & -06:22:29  	   &800         &315& 28 & 6 & 28 \\

37&78, 79&GGD~12-15&                                    06:10:51 & -06:11:54  	   &1000 	&134 & 6 & 6 & 69 \\
38&80&GGD~17&                                           06:12:48 & -06:13:56       &830         &23 & 69 & 69 & 69 \\
39&81&VY~Mon&                                        	06:31:07 &  10:26:05  	   &800 	&23$\pm$5 & 3 & 3 & 4 \\
40&82 to 87&NGC~2264&                                   06:41:05 &  09:40:48  	   &760         & 360 & 66 & 6,30 & 29 \\
41&88&Vela/IRAS~08375-4109&                          	08:39:21 & -41:19:53  	   &700 	&11$\pm$1 &  33 & 33 & 70 \\
42&89&Vela/IRAS~08393-4041&                          	08:41:15 & -40:52:17  	   &700 	& 7$\pm$1 &  33 & 33 & 70 \\
43&90&Vela/IRAS~08404-4033&                          	08:42:17 & -40:44:10  	   &700 	&13$\pm$1 &  33 & 33 & 70 \\
44&91&Vela/IRAS~08448-4343&                          	08:46:35 & -43:54:30  	   &700 	&31$\pm$1 &  33 & 33 & 70 \\
45&92&Vela/IRAS~08470-4243&                          	08:48:48 & -42:54:29  	   &700 	&28$\pm$2 &  33 & 33 & 70 \\
46&93&Vela/IRAS~08470-4321&                          	08:48:48 & -43:32:29  	   &700 	&28$\pm$1 &  33 & 33 & 70 \\

47&94&Vela/IRAS~08476-4306&                          	08:49:26 & -43:17:11  	   &700 	&23$\pm$2 &  33 & 33 & 70 \\
48&95&Vela/IRAS~08477-4359&                          	08:49:33 & -44:10:45  	   &700 	&21$\pm$1 &  33 & 33 & 70 \\
49&96&Vela/IRAS~08500-4254&                          	08:52:18 & -43:05:40  	   &700 	& 7$\pm$1 &  33 & 33 & 70 \\
50&97&Vela Mol. Ridge/BBW192E&                  	08:53:09 & -42:13:03  	   &700-1200 	&32 & 57 & 57 & 57 \\ 
51&98 to 105&Cha~I&                                	11:06:00 & -77:30:00  	   &140         & 180 &  35 & 34  & 35 \\
52&106&Lup~3&                                         	16:09:13 & -39:04:52  	   &150 	&30 & 56 & 56 & 56 \\
53&107 to 115&Ophiuchus&                                16:26:52 & -24:31:42  	   &160         & 337 & 39 & 40, 42 & 42 \\
54&116&MWC~297&                                      	18:27:39 & -03:49:52  	   &450 	&20$\pm$1 & 3 & 3 & 4 \\
55&117&VV~Ser&                                       	18:28:48 &  00:08:40  	   &440 	&17$\pm$5 & 3 & 3 & 4 \\
56&118 to 122&Serpens&                       	        18:29:55 &  01:14:35  	   &310         & 51 & 45 & 44 & 43 \\

57&123, 124, 125&R~Cr~A&                                19:01:54 & -36:57:10  	   &130 	&70 & 6 & 47 & 6 \\
58&126&IRAS~20050+2720&                              	20:07:06 &  27:28:59  	   &700 	&100& 48 & 48 & 48 \\
59&127, 128&BD+40$^o$4124&                              20:20:28 &  41:21:51  	   &980         &33 & 3 & 49 & 49 \\
60&129&S~106&                                        	20:27:27 &  37:22:48  	   &600 	&160& 50 & 50 & 50 \\
61&130&L1228&                                       	20:57:14 &  77:35:47  	   &150 	&48 &  6 & 6 & 6 \\
62&131&L988-e &                                       	21:03:58 &  50:14:38  	   &700 	&45 &  6 & 6 & 6 \\
63&132, 133&LkH$\alpha$ 234/BD+65$^o$1637&              21:43:02 &  66:06:29  	   &1000 	&208 & 6 & 6 & 6 \\
64&134&IC~5146&                                         21:53:29 &  47:16:00       &900-1200 	&800 & 67 & 67 & 67 \\
65&135&LkH$\alpha$~257&                              	21:54:19 &  47:12:11  	   &900 	&6$\pm$6 & 3 & 3 & 4 \\
66&136&Gy~2-21&                                      	22:06:49 &  59:02:11  	   &200 	&$>$5 & 53 & 53 & 53 \\

67&137&S~140&                                        	22:19:18 &  63:18:48  	   &910 	& 16 & 54 & 54 & 54 \\
68&138&S~140N&                                       	22:19:28 &  63:32:56  	   &900 	&12 & 6 & 6 & 6 \\
69&139&L1206&                                       	22:28:52 &  64:13:44  	   &900 	&27 & 6 & 6 & 6 \\
70&140&L1211&                                       	22:47:17 &  62:01:58  	   &750 	&245 & 6 & 6 & 6 \\
71&141&HD~216629&                                    	22:53:15 &  62:08:45  	   &725 	&34$\pm$6 & 3 & 3 & 4 \\
72&142&Cep~A&                                        	22:56:19 &  62:01:58  	   &700 	&580 & 6 & 6 & 6 \\
73&143&Cep~C&                                        	23:05:49 &  62:30:02  	   &750 	&110 & 6 & 6 & 6 \\

\enddata

\end{deluxetable}

\clearpage


\begin{references}
\reference{ada01} Adams, F.~C., \& Myers, P.~C., 2001, \apj, 553, 744
\reference{ali95} Ali, B., \& DePoy, D.~L. 1995, \aj, 109, 709
\reference{all02} Allen, L.~E., Myers, P.~C., Di Francesco, J., Mathieu, R., Chen, H., \& Young, E. 2002, \apj, 566, 993
\reference{alv95} Alves, J.~F., \& Yun, J.~L. 1995, \apj, 438, L107
\reference{asp94} Aspin, C., \& Barsony, M. 1994, \aap, 288, 849
\reference{asp97} Aspin, C., \& Sandell, G. 1997, \mnras, 289, 1
\reference{asp94} Aspin, C., Sandell, G., \& Russell, A.~P.~G. 1994, \aap, 106, 165
\reference{asp90} Aspin, C., \& Walther, D.~M. 1990, \aap, 235, 387
\reference{bar89} Barsony, M., Burton, M.~G., Russell, A.~P.~G., Carlstrom, J.~E., \& Garden, R. 1989, \apj, 346, L93
\reference{bar97} Barsony, M., Kenyon, S.~J., Lada, E.~A., \& Teuben, P.~J. 1997, \apjs, 112, 109
\reference{bei98} Beichman, C.~A., Chester, T.~J., Skrutskie, M., Low, F.~J., \& Gillett, F. 1998, \pasp, 110, 480
\reference{bic03} Bica, E., Dutra, C.~M., \& Barbuy, B., 2003, \aap, 397, 177
\reference{bin87} Binney, J., \& Tremaine, S., ``Galactic dynamics'', Princeton, NJ, Princeton University Press, 1987, p. 187
\reference{bur00} Burkert, A., Stecklum, B., Henning, T., \& Fischer, O. 2000, \aap, 353, 153 
\reference{cam98} Cambresy, L., Copet, E., Epchtein, N., de Batz, B., Borsenberger, J., Fouque, P., Kimeswenger, S., \& Tiphene, D. 1998, \aap, 338, 977
\reference{cam97} Cambresy, L., Epchtein, N., Copet, E., de Batz, B., Kimeswenger, S., Le Bertre, T., Rouan, D., \& Tiphene, D. 1997, \aap, 324, L5
\reference{car00} Carpenter, J.~M., 2000, \aj, 120, 3139
\reference{car+00} Carpenter, J.~M., Heyer, M.~H., \& Snell, R.~L., 2000, \apjs, 130, 381
\reference{car97} Carpenter, J.~M., Meyer, M.~R., Dougados, C., Strom, S.~E., \& Hillenbrand, L.~A. 1997, \aj, 114, 198
\reference{che97} Chen, H., Tafalla, M., Greene, T.~P., Myers, P.~C., \& Wilner, D.~J. 1997, \apj, 475, 163
\reference{che94} Chen, H., \& Tokunaga, A.~T. 1994, \apjs, 90, 149
\reference{cla00} Clarke, C.~J., Bonnell, I.~A., \& Hillenbrand, L.~A., 2000, in Protostars and Planets IV, ed. V. Mannings, A.~P. Boss, \& S.~S. Russell (Tucson: Univ. Arizona Press), 151
\reference{com93} Comeron, F., Rieke, G.~H., Burrows, A., \& Rieke, M.~J. 1993, \apj, 416, 185
\reference{com99} Comeron, F., Rieke, G.~H., \& Neuhauser, R. 1999, \aap, 343, 477
\reference{com96} Comeron, F., Rieke, G.~H., \& Rieke, M.~J. 1996, \apj, 473, 294
\reference{dam87} Dame, T.~M., Ungerechts, H., Cohen, R.~S., de Geus, E.~J., Grenier, I.~A., 
May, J., Murphy, D.~C., Nyman, L.-\AA., \& Thaddeus, P., 1987, \apj, 322, 706
\reference{dep90} DePoy, D.~L., Lada, E.~A., Gatley, I., \& Probst, R. 1990, \apj, 356, L55
\reference{eir92} Eiroa, C., \& Casali, M.~M. 1992, \aap, 262, 468
\reference{elm00} Elmegreen, B.~G., Efremov, Y., Pudritz, R.~E., Zinnecker, H., 2000. Observations and Theory of Star Cluster Formation. In: Protostars and 
Planets IV, Tucson: University of Arizona Press; eds. Mannings, V., Boss, A.~P. and Russell, S.~S., p. 179
\reference{eva89} Evans II, N.~J., Mundy, L.~G., Kutner, M.~L., \& DePoy, D.~L. 1989, \apj, 346, 212
\reference{gio98} Giovannetti, P., Caux, E., Nadeau, D., \& Monin, J.-L. 1998, \aap, 330, 990
\reference{gom93} G\'omez, M., Hartmann, L., Kenyon, S.~J., \& Hewett, R. 1993, \aj, 105, 1927
\reference{gom00} G\'omez, M., \& Kenyon, S.~J. 2000, \aj, 121, 974
\reference{gre92} Greene, T.~P., \& Young, E.~T. 1992, \apj, 395, 516
\reference{hai00} Haisch, K.~E., Lada, E.~A., \& Lada, C.~J. 2000, \aj, 120, 1396
\reference{far01} Hartmann, L.~W. 2001, \aj, 121, 1030
\reference{her02} Herbig, G.~H., \& Dahm, S.~E. 2002, \aj, 123, 304
\reference{hil00} Hillenbrand, L.~A., \& Carpenter, J.~M. 2000, \apj, 540, 236 
\reference{hil98} Hillenbrand, L.~A., \& Hartmann, L.~W. 1998, \apj, 492, 540
\reference{hil95} Hillenbrand, L.~A., Meyer, M.~R., Strom, S.~E., \& Skrutskie, M.~F. 1995, \aj, 109, 280
\reference{hod94} Hodapp, K.-W. 1994, \apjs, 94, 615
\reference{hod93} Hodapp, K.-W., \& Deane, J. 1993, \apjs, 88, 119
\reference{hod91} Hodapp, K.-W., \& Rayner, J. 1991, \aj, 102, 1108
\reference{hod94} Howard, E.~M., Pipher, J.~L., \& Forrest, W.~J. 1994, \apj, 425, 707
\reference{jon79} Jones, B., \& Herbig, G.~H. 1979, \aj, 84, 1872
\reference{jon94} Jones, T.~J., Mergen, J., Odewahn, S., Gehrz, R.~D., Gatley, I., Merrill, K.~M., Probst, R., \& Woodward, C.~E. 1994, \aj, 107, 2120
\reference{kaa99} Kaas, A.~A., 1999, \aj, 118, 558
\reference{ken98} Kenyon, S.~J., Lada, E.~A., \& Barsony, M. 1998, \aj, 115, 252
\reference{lad96} Lada, C.~J., Alves, J., \& Lada, E.~A. 1996, \aj, 111, 1964
\reference{lad03} Lada, C.~J., \& Lada, E.~A., 2003, \araa, 41, 57
\reference{lad00} Lada, C.~J., Muench, A.~A., Haisch, K.~E., Lada, E.~A., Alves, J.~F., Tollestrup, E.~V., \&  Willner, S.~P. 2000, \aj, 120, 3162
\reference{lad93} Lada, C.~J., Young, E.~T., \& Greene, T.~P. 1993, \apj, 408, 471
\reference{lad91} Lada, E.~A., DePoy, D.~L., Evans II, N.~J., \& Gatley, I. 1991, \apj, 371, 171
\reference{lad95} Lada, E.~A., \& Lada, C.~J. 1995, \aj, 109, 1682
\reference{lar95} Larson, R. 1995, \mnras, 272, 213
\reference{li97}  Li, W., Evans II, N.~J., \& Lada, E.~A. 1997, \apj, 488, 277
\reference{lis92} Liseau, R., Lorenzetti, D., Nisini, B., Spinoglio, L., \& Moneti, A. 1992, \aap, 265, 577
\reference{lor93} Lorenzetti, D., Spinoglio, L., \& Liseau, R. 1993, \aap, 275, 489
\reference{luh01} Luhman, K.~L. 2001, \apj, 560, 287
\reference{luh00} Luhman, K.~L., Rieke, G.~H., Young, E.~T., Cotera, A.~S., Chen, H., Rieke, M., Schneider, G., \& Thompson, R.~I. 2000, \apj, 540, 1016 
\reference{mas99} Massi, F., Lorenzetti, D., Giannini, T., Vitali, F., 2000, \aap, 353, 598
\reference{mas99} Massi, F., Giannini, T., Lorenzetti, D., Liseau, R., Moneti, A., \& Andreani, P. 1999, \aaps, 136, 471
\reference{mcc94} McCaughrean, M.~J., \& Stauffer, J.~R. 1994, \aj, 108, 1382
\reference{mey00} Meyer, M.~R., Adams, F.~C., Hillenbrand, L.~A., Carpenter, J.~M., \& Larson, R.~B., 2000. The Stellar Initial Mass function: Constraints 
from Young Clusters and Theoretical Perspectives. In: Protostars and Planets 
IV, Tucson: University of Arizona Press; eds. Mannings, V., Boss, A.~P. and 
Russell, S.~S., p. 121
\reference{mue02} Muench, A.~A., Lada, E.~A., Lada, C.~J. \& Alves, J. 2002, \apj, 573, 366
\reference{nak00} Nakajima, Y., Tamura, M., Oasa, Y., \& Nakajima, T. 2000, \apj, 119, 873 
\reference{oas99} Oasa, Y., Tamura, M., \& Sugitani, K. 1999, \apj, 526, 336
\reference{per99} Persi, P., Marenzi, A.~R., Kaas, A.~A., Olofsson, G., Nordh, L., \& Roth, M. 1999, \aj, 117, 439 
\reference{pet98} Petr, M.~G., du Foresto, V.~C., Beckwith, S.~V.~W., Richichi, A., \& McCaughrean, M.J. 1998, \apj, 500, 825
\reference{reb02} Rebull, L.~M., Makidon, R.~B., Strom, S.~E., Hillenbrand, L.~A., Birmingham, A., Patten, B.~M., Jones, B.~F., Yagi, H., \& Adams, M.~T. 2002, \aj, 123, 1528
\reference{rid03} Ridge, N.~A., Wilson, T.~L., Megeath, S.~T., Allen, L.~E., \& Myers, P.~C., 2003, accepted in \aj   (http://arxiv.org/abs/astro-p/0303401)
\reference{sim99} Simon, M., Close, L.~M., \& Beck, T.~L. 1999, \aj, 117, 1375
\reference{sog97} Sogawa, H., Tamura, M., Gatley, I., \& Merrill, K.~M. 1997, \aj, 113, 1057
\reference{str95} Strom, K.~M., Kepner, J., \& Strom, S.~E. 1995, \apj, 438, 813
\reference{str93} Strom, K.~M., Strom, S.~E., \& Merrill, K.~M. 1993, \apj, 412, 233
\reference{sug91} Sugitani, K., Fukui, Y., \& Ogura, K. 1991, \apjs, 77, 59
\reference{sug95} Sugitani, K., Tamura, M., \& Ogura, K. 1995, \apj, 455, L39
\reference{tap97} Tapia, M., Persi, P., Bohigas, J., \& Ferrari-Toniolo, M. 1997, \aj, 113, 1769
\reference{tej02} Tej, A., Sahu, K.~C., Chandrasekhar, T., \& Ashok, N.~M. 2002, \apj, 578, 523
\reference{tes98} Testi, L., Palla, F., \& Natta, A. 1998, \aaps, 133, 81
\reference{tes99} Testi, L., Palla, F., \& Natta, A. 1999, \aap, 342, 515
\reference{tes97} Testi, L., Palla, F., Prusti, T., Natta, A., \& Maltagliati, S. 1997, \aaps, 320, 159
\reference{tho98} Thompson, R.~I., Corbin, M.~R., Young, E., \& Schneider, G. 1998, \apj, 492, L177
\reference{tut02} Tuthill, P.~G., Monnier, J.~D., Danchi, W.~C., Hale, D.~D.~S., \& Townes, C.~H., 2002, \apj, 577, 826
\reference{wil92} Wilking, B.~A., Greene, T.~P., Lada, C.~J., Meyer, M.~R., \& Young, E.~T. 1992, \apj, 397, 520
\reference{wil97} Wilking, B.~A., McCaughrean, M.~J., Burton, M.~G., Giblin, T., Rayner, J.-T., \& Zinnecker, H. 1997, \aj, 114, 
2029
\reference{yao97} Yao, Y., Hirata, N., Ishii, M., Nagata, T., Ogawa, Y., Sato, S., Watanabe, M., \& Yamashita, T. 1997, \apj,  490, 281
\end{references}
\end{document}